# A hybrid waveguide scheme for silicon-based quantum photonic circuits with quantum light sources

LINGJIE YU,[1] CHENZHI YUAN,[1,2] RENDUO QI,[1] YIDONG HUANG,[1,3,4] AND WEI ZHANG[1,3,4,*]

[1]*Beijing National Research Center for Information Science and Technology (BNRist), Beijing Innovation Center for Future Chips, Electronic Engineering Department, Tsinghua University, Beijing 100084, China.*
[2]*Institute of Fundamental and Frontier Sciences, University of Electronic Science and Technology of China, Chengdu 610054, China*
[3]*Frontier Science Center for Quantum Information, Beijing 100084, China*
[4]*Beijing Academy of Quantum Information Sciences, Beijing 100193, China.*
*\*Email: zwei@tsinghua.edu.cn.*

**Abstract:** We propose a hybrid silicon waveguide scheme to avoid the impact of noise photons induced by pump lights in application scenarios of quantum photonic circuits with quantum light sources. The scheme is composed of strip waveguide and shallow-ridge waveguide structures. It utilizes the difference of biphoton spectra generated by spontaneous four wave mixing (SFWM) in these two waveguides. By proper pumping setting and signal/idler wavelength selection, the generation of desired photon pairs is confined in the strip waveguide. The impact of noise photons generated by SFWM in the shallow-ridge waveguide could be avoided. Hence, the shallow-ridge waveguide could be used to realize various linear operation devices for pump light and quantum state manipulations. The feasibility of this scheme is verified by theoretical analysis and primary experiment. Two applications are proposed and analyzed, showing its great potential on silicon-based quantum photonic circuits.

## 1. Introduction

Quantum photonic circuit has attracted much attention in recent years thanks to its advantages on scalability and stability [1–4]. It is an important way to realize complicated photonic quantum information functions, such as processing units in photonic quantum computation/simulation [5–9] or sender/receiver modules in quantum communication [10–14]. Silicon photonics based on the silicon-on-insulator (SOI) platform is an important route for the quantum photonic circuit. In telecom band, it supports quantum light sources for biphoton state generation [15–18], optical interferometers for pump light manipulation [19] and quantum state manipulation [18,20], and single photon detections for quantum state measurement [21]. Furthermore, it has potential on realizing large-scale systems [4,5]. However, it is still difficult to integrate these functions on one silicon photonic circuit, since they may disturb each other. The impact of pump-light-induced noise photons is one of the problems. Especially, in a circuit integrating quantum light sources based on spontaneous four wave mixing (SFWM) and interferometers for manipulations of pump lights and biphoton states, the pump light for SFWM would propagate through the interferometers and generate unwanted noise photons in them. Several ways have been proposed to solve the problem, such as introducing on-chip pump filters [6,8,22], and utilizing micro-ring resonator to generate photon pairs [22,23]. However, these ways require careful adjustments of filtering/resonating frequencies, which usually depend on phase shifters with electrical controls. As the number of quantum light sources increases with the development of circuit scale, these filters/resonators would increase greatly. It requires more phase shifters with complicated electrical controls, which complicates the circuit.

In this paper, we propose a hybrid silicon waveguide scheme to solve this problem simply in some application scenarios. This scheme is composed of strip waveguides and shallow-ridge waveguides. The biphoton spectrum generated by SFWM in the strip waveguide is much wider than that in the shallow-ridge waveguide. Therefore, if the photon pairs in the frequency region with large spectral difference are selected by post-signal/idler filtering, the selected photon pairs would only come from the strip waveguide, which plays the role of nonlinear medium of SFWM, while the shallow-ridge waveguide can be used to compose various devices for pump light and quantum state manipulation. Hence, it provides an alternative way to avoid the impact of pump-light-induced noise photons, which simplifies the circuit by reducing on-chip filters. In the following sections, firstly, dispersion characteristics and corresponding biphoton spectra in typical strip waveguides and shallow-ridge waveguides are theoretically analyzed. Then, primary experiment is conducted to verify the scheme. Finally, two potential applications are proposed and analyzed theoretically.

## 2. Theoretical analysis

SFWM is frequently used in integrated photon-pair sources. In SFWM, two pump photons are annihilated and two photons called signal/idler photons are generated simultaneously. Energy conservation in SFWM leads to the relation of $\omega_{p1} + \omega_{p2} = \omega_s + \omega_i$, where $\omega_{p1}$, $\omega_{p2}$, $\omega_s$ and $\omega_i$ represent frequencies of the two pump photons and the signal/idler photons, respectively. When frequencies of the two pump photons are different ($\omega_{p1} \neq \omega_{p2}$), the process is called non-degenerate SFWM. If it is stimulated by one mono-color pump light, it is called degenerate SFWM, in which two pump photons have the same frequency ($\omega_{p1} = \omega_{p2}$). The expressions of biphoton spectra generated by the two types of SFWM are provided in Appendix, by which we can analyze the difference of the photon pair generation in strip waveguides and shallow-ridge waveguides, showing the principle of the proposed scheme.

Fig. 1(a) and 1(b) show two typical designs of 220-nm-high strip and shallow-ridge waveguides with $SiO_2$ cladding. The strip waveguide has a core width of 500 nm, which supports the fundamental quasi-transverse electric (quasi-TE) mode. The shallow-ridge waveguide has a ridge with a width of 1 μm and a height of 70 nm, respectively. It also supports its fundamental quasi-TE mode. Theoretical analysis shows that fundamental quasi-TE modes are the lowest modes in these two types of waveguides. Hence, the light can smoothly transit between the two types of waveguides by a proper adiabatic transition structure [24,25]. It is worth noting that the strip waveguide and shallow-ridge waveguide all support fundamental quasi-TM modes. However, it has been proven that SFWM process in fundamental quasi-TM modes is far weaker than that in fundamental quasi-TE mode [26]. On the other hand, it is difficult for vector SFWMs between the quasi-TE and quasi-TM modes to realize phase matching in telecom band due to the large difference on their phase constants [17]. As a result, the fundamental quasi-TE mode in strip waveguides is the best choice to realize quantum light sources on silicon photonic quantum circuit. Hence, only the fundamental quasi-TE modes in these two waveguides are considered in this work. Their dispersion characteristics are calculated numerically and shown in Fig. 1(c). It can be seen that the shallow-ridge waveguide shows a large normal dispersion in telecom band, which is mainly due to the material dispersion of Silicon. On the other hand, the strip waveguide shows the feature of a small anomalous dispersion. It is due to the strong mode field confinement in the strip waveguide, which introduces a large anomalous waveguide dispersion to compensate for the material dispersion. It can be seen that the dispersion characteristics of the two waveguides satisfy the requirement of the proposed scheme. The effective nonlinear coefficients (γ) of the fundamental quasi-TE modes in two waveguides are also calculated numerically. It shows that the strip waveguide has a much higher γ (223.3 m⁻

$^{-1}W^{-1}$ at 1552.5 nm) than that of the shallow-ridge waveguide (93.5 m$^{-1}$W$^{-1}$ at 1552.5 nm), since the strip waveguide has a much smaller effective mode field area.

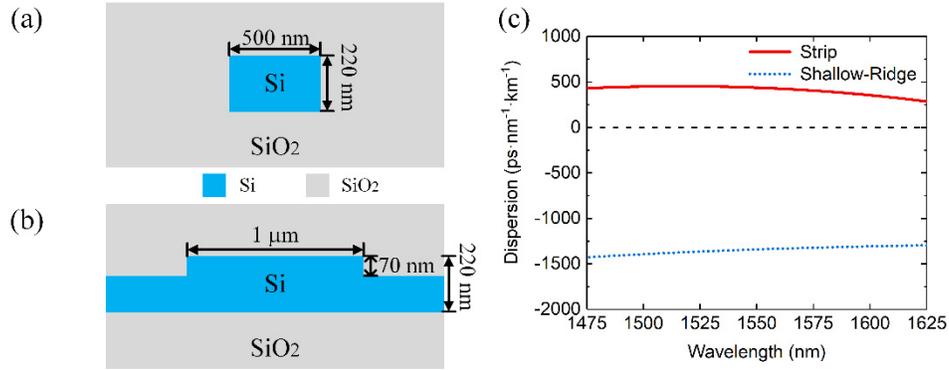

Fig. 1. Typical structures of (a) a strip waveguide and (b) a shallow-ridge waveguide, with waveguide parameters close to the fabricated samples in the experiment. (c) The calculated dispersions of the two waveguides in telecom band. The strip waveguide has small anomalous dispersion, while the shallow-ridge waveguide has large normal dispersion.

Based on the calculated dispersions and effective nonlinear coefficients, the phase mismatch and the biphoton spectra of SFWM in these two waveguides can be calculated. In the calculation, the length of the strip waveguide is 5 mm. Three shallow-ridge waveguide lengths are considered for comparison, which are 3 mm, 8 mm and 15 mm, respectively. In the case of degenerate SFWM, a mono-color pulsed pump light at 1552.5 nm with a peak power of 1 W is used. Fig. 2(a) shows the phase mismatch of the strip and shallow-ridge waveguide under the peak pump power. It can be seen that the phase mismatch of the shallow-ridge waveguide grows faster than that of the strip waveguide when the absolute value of frequency detuning increases. It leads to the difference of biphoton spectra, which is shown in Fig. 2(b). It shows that the shallow-ridge waveguides with different lengths all have narrow biphoton spectra with similar bandwidths, which are close to the pump wavelength. On the other hand, the biphoton spectrum in the 5-mm strip waveguide is much wider than those in the shallow-ridge waveguides. It can be expected that if the frequencies of optical filters in the post-selection are set at large frequency detuning, only photon pairs generated in the strip waveguide will be selected. Figure 2(c) and 2(d) shows the calculated phase mismatch and biphoton spectra in the cased of non-degenerate SFWM, respectively. Two continuous-wave (CW) pump lights in the calculation are set at 1528 nm and 1582 nm, respectively, with the same power of 10 mW. The waveguide parameters in the calculation are the same as those in Fig. 2(a) and 2(b). It also can be seen that the phase mismatch of the shallow-ridge waveguide grows faster than that of the strip waveguide when the absolute value of frequency detuning increases. Hence, the biphoton spectra of shallow-ridge waveguides are close to the pump wavelengths and that of the strip waveguide is much wider than them. The large difference appears both in the region between the two pump wavelengths and the region outside the two pump wavelengths. Both regions are suitable to set the frequencies of signal/idler filters in the post-selection. Particularly, if the center frequencies of the signal/idler filters both select the average frequency of the two pump lights, photon pairs with identical photons would be post-selected.

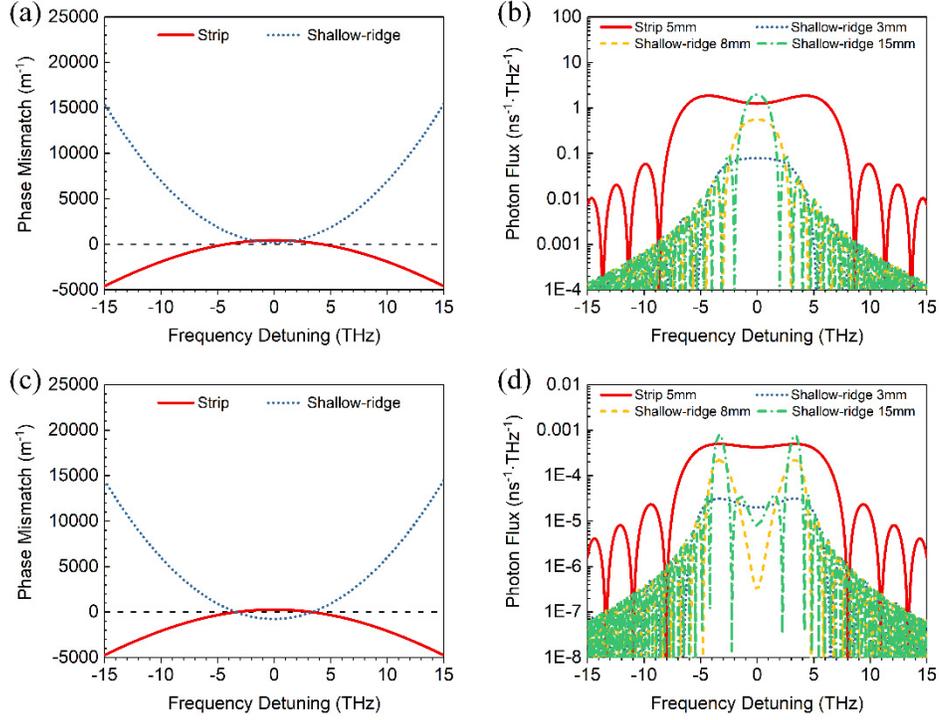

Fig. 2. The calculated (a) phase mismatch and (b) biphoton spectra of degenerate SFWM processes in the strip waveguide and the shallow-ridge waveguides. The calculated (c) phase mismatch and (d) biphoton spectra of non-degenerate SFWM processes in the strip waveguide and the shallow-ridge waveguides. The length of the strip waveguide is 5 mm, while three shallow-ridge waveguide lengths are considered, which are 3 mm, 8 mm and 15 mm, respectively.

## 3. Experiment

The principle of the proposed hybrid waveguide scheme is verified experimentally. The biphoton spectra of non-degenerate SFWM in hybrid waveguide samples are measured and compared with those in the shallow-ridge waveguide samples under the same waveguide lengths and pump conditions.

The hybrid waveguide samples are fabricated on the 220 nm SOI substrate by electron beam lithography and inductively coupled plasma etching. The parameters of the strip waveguide and the shallow-ridge waveguide are the same as those for theoretical analysis in the preceding section, which have been shown in Fig. 1(a) and 1(b), respectively. Two kinds of hybrid waveguide samples are prepared. In one of them, the lengths of strip waveguide and shallow-ridge waveguide are 3 mm and 5 mm, respectively. In the other one, the length of strip waveguide is also 3mm, but the length of shallow-ridge waveguide increases to 14 mm. Hence, the total lengths of the two kinds of samples are 8 mm and 17 mm, respectively. Between the strip waveguide and the shallow-ridge waveguide, a transition section of 100 microns in length is designed. Figure 3(a) shows the sketch of the transition section. It is a taper structure which could be fabricated by two etching processes. In the sketch, the regions with different colors indicate different etch depths. The structure parameters are indicated in the sketch. A scanning electronic microscope (SEM) picture shown in Fig. 3(b) corresponds to a part of the transition section in the sketch, indicating the actual fabrication result on the samples. Since it supports adiabatic transition between the fundamental quasi-TE modes in these two waveguides, its loss is quite small. On the same chip, several shallow-ridge waveguides with lengths of 8 mm and 17 mm are also fabricated for comparison. Figure 3(c) shows the sketches of the hybrid and shallow-ridge waveguides. Grating structures are fabricated at the two ends of these waveguides for fiber coupling. Figure 3(d) and 3(e) show the typical measured transmission spectra of 8-mm and 17-mm waveguide samples, respectively. The total insertion losses of the waveguide samples are mainly from the coupling loss of grating couplers and the propagation losses of waveguides. The coupling loss spectra of the grating couplers show filtering effect, with a minimum coupling loss of about 4.5 dB and a 3 dB bandwidth over 50 nm. According to the measurement result shown in Fig. 3, the attenuation coefficients of strip waveguide and shallow-ridge waveguide could be estimated to be ~3 dB/cm and ~10 dB/cm, respectively. It can be seen that the insertion losses of the hybrid waveguide are a little higher than those of the shallow-ridge waveguide. It is mainly due to the attenuation difference between the strip waveguide and shallow-ridge waveguide.

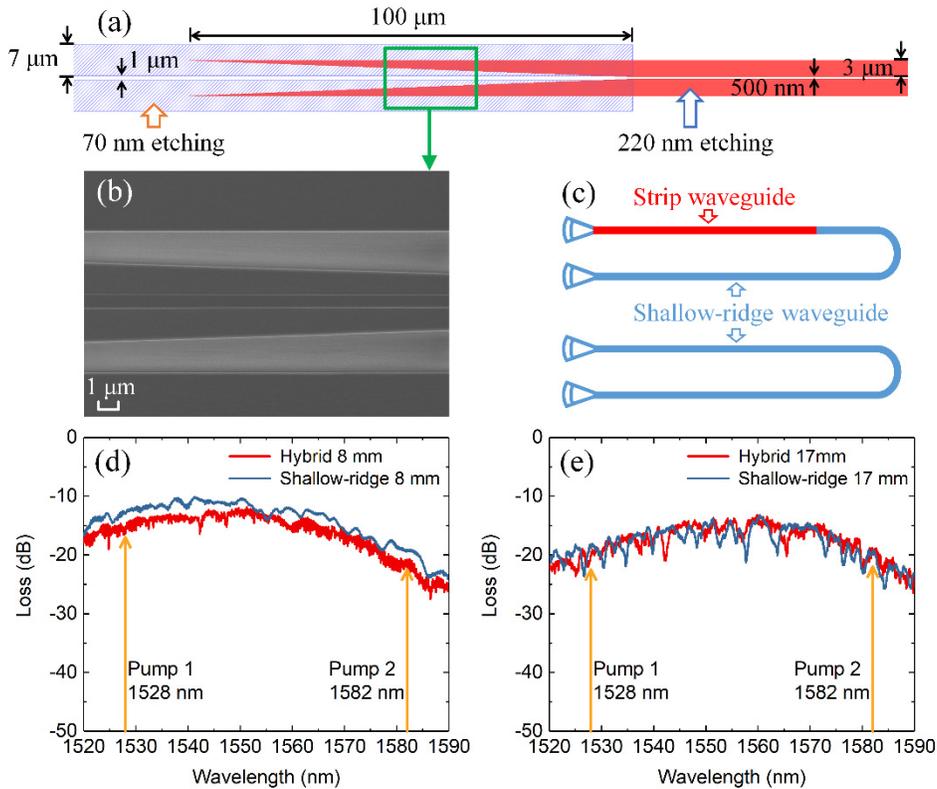

Fig. 3. The waveguide samples and their typical measured transmission spectra. (a) The sketch of the transition section in the hybrid waveguide sample with parameters indicated and (b) a SEM picture of a part of the section. (c) The sketches of the hybrid and shallow-ridge waveguides on the chip. Typical measured transmission spectra of (d) 8-mm waveguide samples and (e) 17-mm waveguide samples.

According to Fig. 2(d), the preferred frequency region of signal/idler photons is near the average frequency of the two pump lights in the case of non-degenerate SFWM. In this region, it is easy to separate signal and idler photons by commercial components for optical communications. The experiment setup is shown in Fig. 4. Two tunable continuous-wave (CW) lasers generate pump lights at wavelengths of 1528 nm and 1582 nm, respectively. Two optical filters made of coarse wavelength division multiplexing (CWDM) devices are used to combine them and suppress the noise photons in the frequency region of signal/idler photon collection. Then the pump lights are coupled into the fundamental quasi-TE modes of the waveguide samples by an optical fiber. Two fiber polarization controllers (FPC) are used to adjust the pump light polarizations. When the pump lights propagate along the waveguide, photon pairs are generated by non-degenerate SFWM. They are coupled out by another optical fiber. Then optical filters made of dense wavelength division multiplexing (DWDM) devices are used to select signal and idler photons at specific frequencies. In the experiment, four filter groups are used for different signal/idler wavelength selections, which is shown in Table 1. The extinction ratios of all the filters are higher than 120 dB by cascading several DWDM components. The selected signal and idler photons are detected by two single photon detectors (SPDs, Id 220, IDQ Inc.) and recorded by a time correlated single photon counting circuit (TCSPC, DPC-230, Becker & Hickl GmbH).

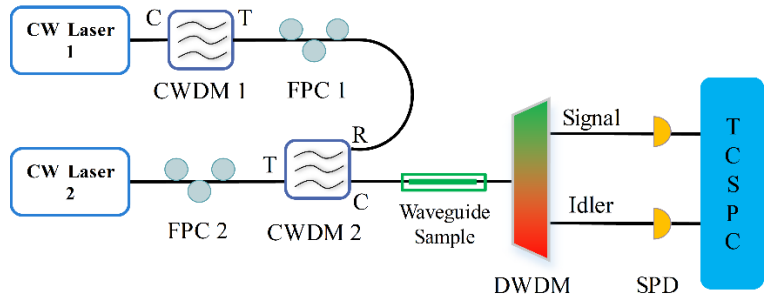

Fig. 4. Experiment setup for photon pair generation by non-degenerate SFWM. CW Laser 1/2: two tunable CW lasers for pump light generation; CWDM 1/2: optical filters made of coarse wavelength division multiplexing devices, C: com, T: transmission, R: reflection; FPC: fiber polarization controller; DWDM: optical filter made of dense wavelength division multiplexing devices; SPD: single photon detector; TCSPC: time correlated single photon counting circuit.

Table 1. Optical filters for signal/idler photons

| Filter Group Number | ITU Channel | Center wavelength /nm | Loss /dB | 3 dB bandwidth /nm | ITU Channel | Center wavelength /nm | Loss /dB | 3 dB bandwidth /nm |
|---|---|---|---|---|---|---|---|---|
| 1 | C30 | 1553.3 | 2.9 | 0.47 | C27 | 1555.7 | 2.4 | 0.59 |
| 2 | C33 | 1550.9 | 3.1 | 0.43 | C24 | 1558.2 | 1.9 | 0.45 |
| 3 | C35 | 1549.3 | 2.7 | 0.5 | C22 | 1559.8 | 2.9 | 0.47 |
| 4 | C46 | 1540.6 | 3.3 | 0.52 | C11 | 1568.8 | 2.0 | 0.51 |

In the experiment, the signal/idler wavelengths are selected near the center wavelength of operation band of the grating coupler. The two pump lights have a large wavelength space. Hence, their insertion losses are a little higher due to the filtering effect of the grating coupler. However, it can be compensated by pump power enhancement easily. The two arrows shown in Fig. 3(d) and 3(e) indicate the pump wavelengths in the experiment (1528 nm and 1582 nm). The insertion losses of the 8-mm hybrid waveguide at the two wavelengths are 15 dB and 21 dB, respectively. Those of the 8-mm shallow-ridge waveguide are 13 dB and 19 dB, respectively. At the same pump wavelengths, the insertion losses are 20 dB and 19 dB for the 17-mm hybrid waveguide, 18 dB and 18.5 dB for the 17-mm shallow-ridge waveguide, respectively. It can be seen that for the two 8-mm waveguide samples, the insertion loss differences between the two pump lights are all about 6 dB, which is contributed by the filtering effect of two grating couplers. Hence, the pump powers of the pump two lights are set as 7.5 dBm at 1528 nm and 11.2 dBm at 1582 nm for the 8-mm samples, respectively, to make the coupled powers of the two pump lights close. On the other hand, for the two 17-mm waveguide samples, the insertion loss differences between the two pump lights are close.

Hence, the pump powers of the pump two lights are set as 10.5 dBm at 1528 nm and 11.2 dBm at 1582 nm, respectively.

Figure 5(a) is a typical coincidence measurement result for the photon pair generated in the 8-mm hybrid waveguide. In this measurement, the signal and idler wavelengths are selected at 1553.3 nm and 1555.7 nm, respectively, which are close to the average wavelength of two pump lights. The single side count rates of signal and idler photons are 11.6 kHz and 15.0 kHz, respectively. The corresponding result of the 8-mm shallow-ridge waveguide under the same pump level is shown in Fig. 5(b). It can be seen that the histogram of hybrid waveguide shows a distinct coincidence peak compared to that of the shallow-ridge waveguide sample. Hence, the photon pairs generated in the hybrid waveguide sample are mainly from the strip waveguide.

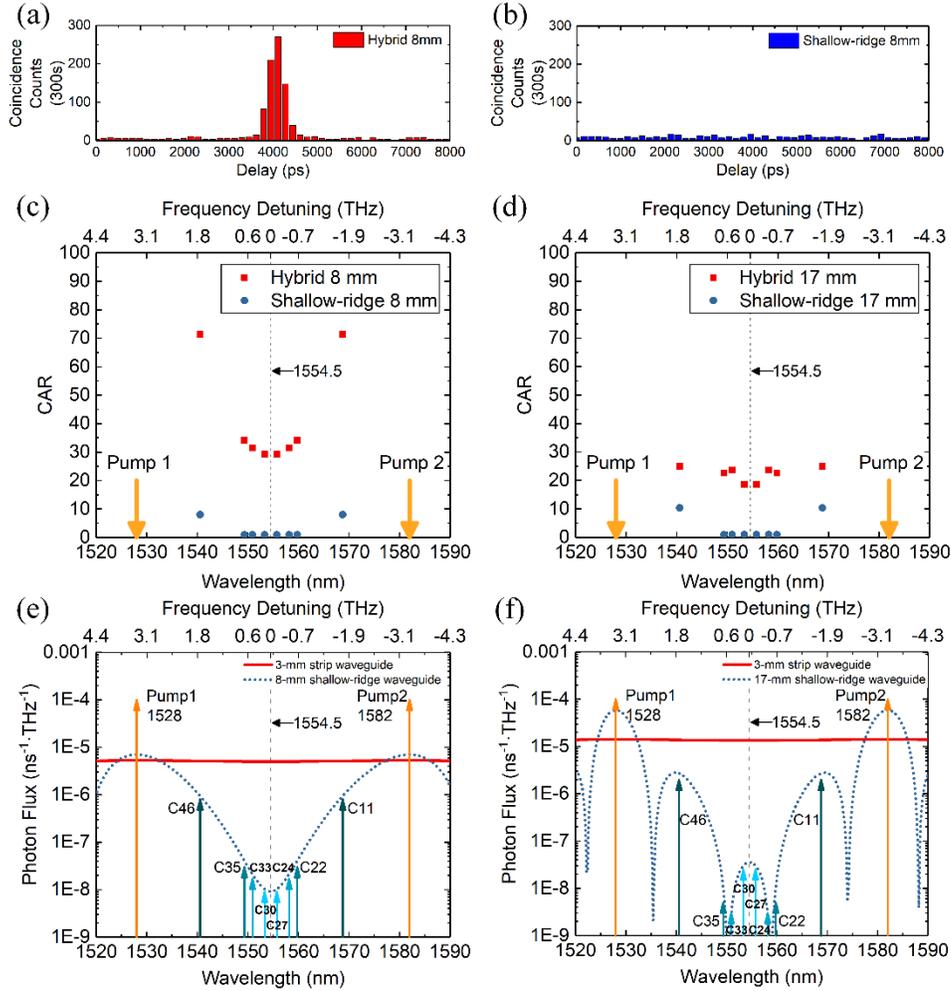

Fig. 5. Experimental results of photon pair generation in the waveguide samples. Typical results of the coincidence measurement of 8-mm (a) hybrid and (b) shallow-ridge waveguides; coincidence-to-accidental ratios (CAR) comparison of (c) 8-mm and (d) 17-mm waveguide samples under different signal/idler wavelengths. (e) Calculated SFWM photon flux spectra of the in the 3 mm strip waveguide in the 8-mm hybrid waveguide sample and 8-mm shallow-ridge waveguide sample, respectively, under the pump condition used in the experiment. (f) Calculated SFWM photon flux spectra of the in the 3 mm strip waveguide in the 17-mm hybrid waveguide sample and 17-mm shallow-ridge waveguide sample, respectively, under the pump condition used in the experiment. The orange arrows indicate the pump wavelengths while the blue arrows indicate the center wavelengths of the signal/idler filters used in the experiment. In (c), (d), (e) and (f), the bottom X axis indicates the wavelength while the top X axis indicates the corresponding frequency detuning relative to 1554.5 nm, which is the average wavelength of two pump lights.

Based on the coincidence measurement results, coincidence-to-accidental ratios (CAR) can be calculated to indicate the photon pair generation. In Fig. 5(c), the red squares show the measured CARs of photon pairs generated in the 8-mm hybrid waveguide under different signal/idler wavelength selections and the same pump condition. The coincidence counts are calculated based on the average counts of 5 time bins covering the coincidence peak. The accidental coincidence counts are the average counts of time bins outside the coincidence peak. The results of the shallow-ridge waveguide sample under the same pump condition are shown as blue circles for comparison. It can be seen that in the region near the average wavelength of two pump lights (10 nm in width), the CARs of the photon pairs generated in the hybrid waveguide are about 30, while the CARs of the photon pairs generated in shallow-ridge waveguide are close to 1. It shows that there are almost no photon pairs generated in this wavelength region from the shallow-ridge waveguide in the hybrid sample. When signal and idler photons are selected with larger wavelength space, such as 1540.6 nm and 1568.8 nm shown in Fig. 5(c), the CAR of the photon pairs generated in the hybrid waveguide rises, while that of the shallow-ridge waveguide is also higher than 1. It shows that part of the photon pairs at these wavelengths are from the shallow-ridge waveguide in the hybrid sample, which may disturb the biphoton state generated in the strip waveguide. The CARs of the photon pairs generated in 17-mm waveguide samples are also measured. The results are plotted in Fig. 5(d), showing similar properties with those in Fig. 5(c). Similar with the theoretical analysis in Section 2, we calculate the photon flux spectra of the SFWM in the 3-mm strip waveguide, 8-mm shallow-ridge waveguide and 17-mm shallow-ridge waveguide, respectively, under the pump condition used in the experiment, which are plotted in Fig 5(e) and 5(f), respectively. It shows that the photons generated in the shallow-ridge waveguide would increase obviously when the center wavelengths of the signal/idler filters are selected at 1540.6 nm and 1568.8 nm, respectively. This primary experiment verifies that in the hybrid waveguide scheme, the photon pair generation and the manipulation of pump light and quantum state can be realized separately by proper post-selections of signal/idler photons.

It is worth noting that the propagation performance of the waveguide samples used in the experiment is limited by our fabrication conditions. It can be expected that the propagation losses of the two types of waveguides could be reduced by improving the fabrication. Small attenuation coefficients of 0.27dB/cm [24,27] and 2 dB/cm [28] have been reported for the shallow-ridge and strip waveguides, respectively. Reduction of the waveguide propagation loss indicates that more pump light would be leaked to the quantum state manipulation part, leading to a more serious situation on noise photon generation in this part. The proposed scheme would provide a simple way to solve this problem effectively.

## 4. Simulation of potential applications

To show the potential applications of the proposed hybrid waveguide scheme, we provide two examples of silicon-based quantum photonic circuits and show the effect of the scheme by numerical simulation. The target photon pairs are generated in the two examples by degenerate SFWM and non-degenerate SFWM, respectively. The parameters of the strip waveguide and shallow-ridge waveguide are the same as those illustrated in Fig. 1(a) and 1(b), respectively. The pump conditions in the simulation are also the same as those used in the calculation of Fig. 2.

*4.1 The circuit for time-bin entanglement generation by degenerate SFWM*

The sketch of the circuit design and its measurement setup is shown in Fig. 6(a). The circuit includes an unbalanced Mach-Zehnder interferometer (UMZI) and a piece of silicon waveguide as the nonlinear medium for degenerate SFWM. The pulsed pump light is injected into the circuit by a grating coupler. Then it propagates through the UMZI, by which a pump pulse splits into two cascaded ones with a phase difference of $\alpha$. (The orange part represents the thermal-optic phase shifter controlling $\alpha$.) Then the pump pulses pass through the silicon waveguide and generate time-bin entangled biphoton state of

$$|\xi\rangle = \frac{1}{\sqrt{2}}(|0,0\rangle + e^{2i\alpha}|1,1\rangle) \tag{1}$$

by SFWM, where 0 and 1 represents the front and the rear time bins, respectively. Since the pulsed pump light pass through the UMZI before the nonlinear medium, it can be expected that the SFWM in the UMZI would disturb the final biphoton state. An additional on-chip filter would be required between them. To simplify the circuit design, we use the hybrid waveguide scheme to avoid the impact of nonlinear effect in the UMZI. In Fig. 6(a), the grey rectangle with black frame represents the chip, on which the strip and shallow-ridge waveguides are indicated by red and blue lines, respectively. Photon pairs are expected to be generated in the strip waveguide by the degenerate SFWM process, which is realized by proper post-selection using optical filters before single photon detectors.

In simulation, the center wavelength of the pulsed pump light is set at 1552.5 nm. A square pump pulse profile is assumed with a peak power of 1 W. To show the contributions of photon pair generation at different parts on the chip, the photon fluxes of SFWM in the long/short arms of the UMZI and the strip waveguide are calculated under the peak power of 0.5 W, 0.5 W and 0.25 W, respectively, which are determined by the beam splitters in the UMZI. The photon fluxes of the long and short arms also have an additional 3 dB loss, introduced by the second beam splitter in the UMZI. The calculation only takes waveguide dispersion, effective nonlinear coefficient and waveguide length into consideration. The attenuation and coupling loss of the waveguide are not considered. The lengths of the waveguides in each stage are shown in Fig. 6(a). The length difference between the long and the short arms in the UMZI is 11.5 mm. In telecom band, the effective refractive index of the shallow-ridge waveguide is about 2.6. The corresponding time difference between the cascaded two pump pulses after the UMZI is 100 ps.

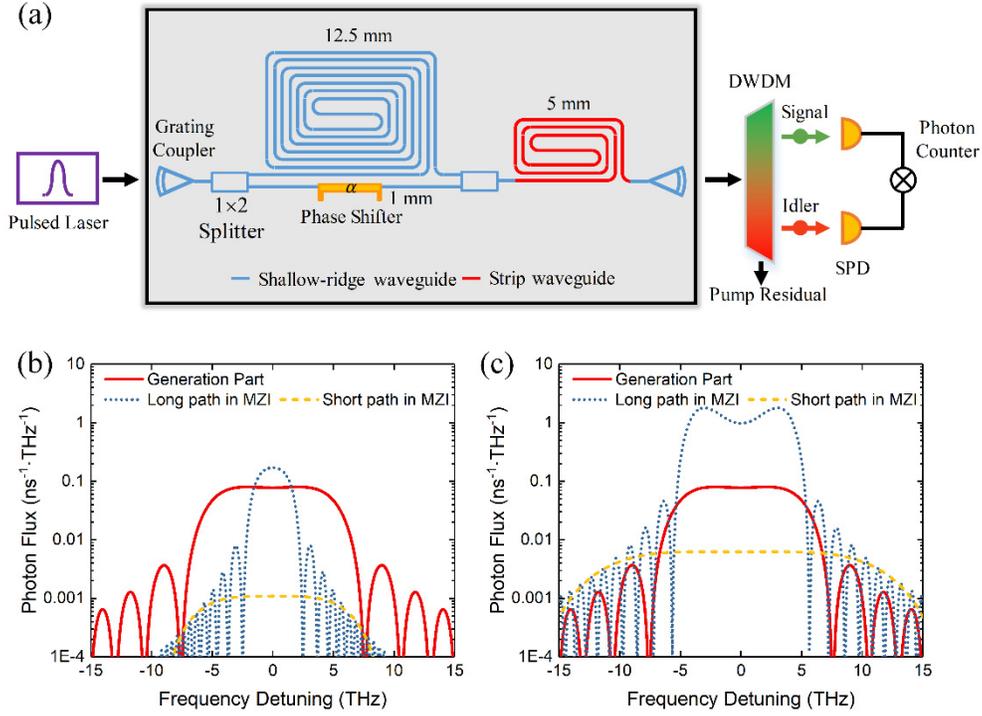

Fig. 6. (a) The sketch of the circuit for time-bin entanglement generation and its measurement setup. The grey rectangle with black frame represents the chip. The blue line indicates the part of shallow-ridge waveguide, which realize the UMZI and the grating couplers. The red line indicates the strip waveguide, which is used as the nonlinear medium for degenerate SFWM. By the optical filter system before SPDs, the post-selected photon pairs are mainly generated in the strip waveguide and the impact of SFWM in the UMZI can be avoided. (b) The calculated biphoton spectra of the degenerate SFWM in the strip waveguide and the shallow-ridge waveguide in the circuit. (c) The calculated biphoton spectra of the degenerate SFWM in different parts if the UMZI is also realized by the strip waveguide.

Figure 6 (b) shows the calculation results. It can be seen that the biphoton spectrum of the strip waveguide is quite broad. On the other hand, the biphoton spectra of the long and short arms in the UMZI are narrow, due to the dispersion property of the shallow-ridge waveguide. If photon pairs with large frequency detuning are selected (2.5~5 THz), it can be expected that the contribution of the strip waveguide is at least 10 times larger than that of the shallow-ridge waveguide. For comparison, this circuit design is also analyzed under the condition that all the waveguides are changed to strip waveguides. The calculation results are shown in Fig. 6(c). It can be seen that the biphoton spectral bandwidth of the long path in the UMZI is similar with that of the strip waveguide for SFWM, and its intensity is even higher. Hence, the effect of the SFWM in the UMZI part cannot be avoid in this case. This comparison shows that the hybrid waveguide scheme is suitable for this circuit of on-chip time-bin entanglement generation. It is also effective for other quantum photonic circuits with pump light manipulation parts.

### *4.2 The circuit for path entanglement generation by non-degenerate SFWM with entanglement distribution and analysis*

The sketch of the circuit design and its measurement setup is shown in Fig. 7(a). In this design, photon pairs are generated by non-degenerate SFWM. Signal and idler photons are frequency-degenerate and their frequency is selected at the average frequency of the two pump lights. It is realized by the optical filters before the SPDs. The circuit can be divided into three parts: quantum light source, quantum state distribution and quantum state analyzer, which are separated by two dashed lines in Fig. 7(a).

In the part of quantum light source, two pump lights with different frequencies are coupled into the circuit by two grating couplers. A 50:50 beam splitter is used to combine the two pump lights and then equally divide them into two Mach-Zehnder interferometers (MZIs, indicated by MZI A/B in Fig. 7(a)). Each of the two MZIs is composed of two 50:50 beam splitters and two 5-mm strip waveguides. In each MZI, the 5-mm strip waveguides are used as the nonlinear media of non-degenerate SFWM to generate photon pairs, and the biphoton interference occurs when the generated photon pairs are sent to the output beam splitter. Taking MZI A as an example, for the selected frequency-degenerate photon pairs, the biphoton state generated in the two 5-mm strip waveguides is

$$|\eta_A\rangle = \frac{1}{\sqrt{2}}\left(|2,0\rangle - e^{i2\theta_A}|0,2\rangle\right) \quad (2)$$

In the two kets of the expression, 0 and 2 represent the photon number at the upper or lower arms of the MZI A. $2\theta_A$ is the relative phase difference between the two kets, which can be adjusted by the thermal-optic phase shifter (PS2-A in Fig. 7(a)) on one of the strip waveguides. The biphoton interference at the output beam splitter of MZI A transfers the biphoton state to the superposition of the bunch state and the anti-bunch state.

$$|\zeta_A\rangle = \frac{1+e^{2i\theta_A}}{2\sqrt{2}}\left(-|2,0\rangle + |0,2\rangle\right) + \frac{i(1-e^{2i\theta_A})}{2}|1,1\rangle \quad (3)$$

By adjusting the relative phase difference $2\theta_A$, the bunch state part in Eq. (3) could be removed and only the anti-bunch state part is left, i.e., $|\zeta_A\rangle = |1,1\rangle$. For MZI B, the processes of photon pair generation and biphoton interference are the same as those in MZI A. Hence, at the output ports of MZI B, we can also get the biphoton state of $|\zeta_B\rangle = |1,1\rangle$ by adjusting the thermal-optic phase shifter in MZI B (PS2-B in Fig. 7(a)). Under this condition, the generated biphoton state $|\psi\rangle$ by the quantum light source is path entangled, which can be expressed as

$$|\psi\rangle = \frac{1}{\sqrt{2}}\left(|A_1, A_2\rangle + e^{2i\alpha}|B_1, B_2\rangle\right) \quad (4)$$

Where $A_1/A_2$ and $B_1/B_2$ indicate the output ports of MZI A and MZI B, respectively (shown in Fig. 7(a)). $2\alpha$ is the phase difference between the two kets, which can be controlled by the phase shifter PS1 shown in Fig. 7(a).

The part of quantum state distribution includes four waveguides, which are used to distribute the path entangled photon pairs to different places on the chip. In the calculation, the length of each waveguide is 7 mm. The photon pairs are finally sent to the part of quantum state analyzer to measure the property of path entanglement. The analyzer consists of two balanced MZIs, which can realize the $\hat{R}_z$ and $\hat{R}_y$ rotations on the signal/idler photons [23,29]. At the outputs of the MZIs, four grating couplers are used to couple the photons to optical fibers. Optical filters are used to select frequency-degenerate photon pairs and remove unwanted photons and residual pump light before the SPDs.

In this circuit design, hybrid waveguide scheme is used to confine the photon pair generation processes into the part of quantum light source. The strip and shallow-ridge waveguides are indicated by red and blue lines, respectively. The orange part represents the thermal-optic phase shifters. It can be seen that the strip waveguides are only used in the quantum light source part as nonlinear media. Other waveguides on the chip are all shallow-ridge waveguides. In the simulation, the two pump lights are set at wavelengths of 1528 nm and 1582 nm, respectively, both with the power of 10 mW. To show the contributions of photon pair generation in different parts on the chip, the photon fluxes of SFWM in different parts are calculated, including the strip waveguides in the quantum light sources, the four waveguides for quantum state distribution and the waveguides of the MZIs in the quantum state analyzer part. The calculation only takes waveguide dispersion, effective nonlinear coefficient and waveguide length into consideration. Waveguide attenuation and the loss of fiber coupling are not considered. The splitting ratios of all the on-chip beam splitters are 50:50. The lengths of the waveguides in each part are shown in Fig. 7(a). In the calculation, each of the two pump lights has the power of 2.5 mW in all the waveguides. This pump condition is under the assumption that the pump lights distribute uniformly among the waveguides in each part, which provides a good estimation of the photon pair generation by non-degenerate SFWM in each part.

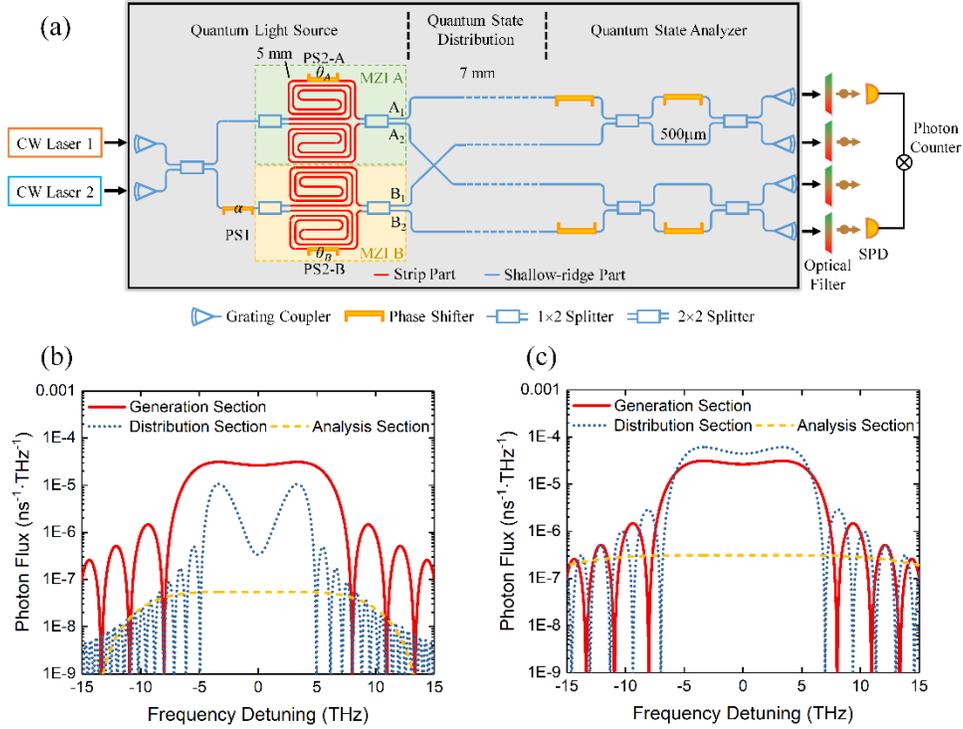

Fig. 7 (a) the sketch of the circuit for path entanglement generation, distribution and analysis with its measurement setup. The grey rectangle with black frame represents the chip. The two black dashed lines divide the chip into three parts: the quantum light sources, the quantum state distribution and the quantum state analyzer. The red lines indicate the 5-mm strip waveguides for photon pair generation by non-degenerate SFWM. The blue lines indicate the shallow-ridge waveguides. The phase shifters are shown in orange. By the optical filter system before SPDs, the frequency-degenerate photon pairs are selected and the impact of non-degenerate SFWM in the parts of quantum state distribution and state analyzer can be avoided. (b) The calculation results of the biphoton spectra of different parts, including the strip waveguides in quantum light sources, the four waveguides for quantum state distribution and the waveguides in the MZIs in the quantum state analyzer. (c) The calculation results if all the waveguides in calculation are realized by strip waveguides.

Fig. 7(b) shows the biphoton spectra of waveguides in different parts. The frequency detunings of the two pump lights are ±3.3 THz, respectively. It can be seen that the biphoton intensity in the quantum light sources is at least 10 times higher than that in the distribution part or analyzer part at zero frequency detuning. For comparison, this circuit design is also analyzed under the condition that all the waveguides are changed to strip waveguides. The calculation results are shown in Fig. 7(c). It can be seen that the biphoton intensity in the distribution part is higher than that in quantum light sources at zero frequency detuning. Hence, the impact of the SFWM in the distribution part cannot be avoided in this case. This comparison shows that the hybrid waveguide scheme is suitable for this circuit, by which the nonlinear effect in quantum state manipulation part can be avoided without on-chip filters. It is also effective for other quantum photonic circuits with quantum state manipulation parts.

It is worth noting that the waveguide attenuation is not considered in the simulations of both applications for simplification. If the waveguide attenuation is considered, it could be expected that the pump powers should be enhanced to compensate for the waveguide loss in the pump light manipulation part to support the same photon pair generation rate in the strip waveguides. However, since this part is realized by shallow-ridge waveguides with low attenuation, the pump power enhancement is limited if the waveguides in this part are not too long. As a result, the increase of the noise photon generation in this part is also limited. On the other hand, the strip waveguide for the photon pair generation is usually several millimeters in length, which has a relatively high attenuation coefficient. It would reduce the powers of the pump lights propagating through the quantum state manipulation part, leading to an obvious reduction of the noise photon generation in this part. The waveguide attenuation also impacts the pump level required by the on-chip quantum light sources. The effective length of nonlinear processes in a waveguide is limited by its attenuation. Hence, long strip waveguides with low attenuation are preferred in the on-chip quantum light sources, since it is helpful to reduce the pump level by extending the effective length of nonlinear processes.

## 5. Conclusion

A hybrid silicon waveguide scheme is proposed to solve the problem of pump light induced noise in quantum photonic circuits with quantum light sources based on SFWM. The scheme is composed of strip waveguide and shallow-ridge waveguide structures. It utilizes the difference of biphoton spectra generated in these two types of waveguides. By proper pumping setting and signal/idler wavelength selection, the impact of noise photons generated by SFWM in shallow-ridge waveguides can be avoided. The photon pairs finally detected are generated only in the strip waveguide part. Since the part of the shallow-ridge waveguide has no contribution on the detected photons, it could be used to realize various linear optical devices for pump light manipulation and quantum state manipulation. In this paper, waveguide dispersion and corresponding biphoton spectra are theoretically analyzed for the two type of waveguides, showing the principle of the proposed scheme. It is also verified by a primary experiment. Two potential applications of this scheme on quantum photonic circuits are proposed and analyzed theoretically, showing its feasibility on silicon-based quantum photonic circuits with quantum light sources.

**Funding**

This work was supported by the National Key R&D Program of China under Contract No. 2017YFA0303704 and 2018YFB2200400; the National Natural Science Foundation of China under Contract No. 61575102, 61875101, 91750206 and 61621064; Beijing National Science Foundation under Contract No. Z180012; Beijing Academy of Quantum Information Sciences under Contract No. Y18G26.

**Disclosures**

The authors declare no conflicts of interest.

**Appendix: Expressions of the biphoton spectra generated by non-degenerate and degenerate SFWM processes in nonlinear waveguides**

For SFWM in nonlinear waveguides, phase mismatch $\Delta k$ plays an important role, which deeply influences the spectrum of photon pair generation [30]. It can be expressed as

$$\Delta k = \Delta k_{NL} + \Delta k_{L} \tag{5}$$

Where $\Delta k_{NL}$ is the nonlinear phase mismatch of the waveguide, which will be discussed later, and $\Delta k_{L}$ is the linear phase mismatch of the waveguide, which can be expressed as,

$$\Delta k_{L} = -k_{P1} - k_{P2} + k_{s} + k_{i} \tag{6}$$

Where $k_{P1}$, $k_{P2}$, $k_{s}$ and $k_{i}$ are propagation constants of the two pump, signal and idler photons, respectively. The linear phase mismatch can be expanded near the average of the two pump frequencies as Eq. (7), which only has even order terms [31].

$$\Delta k_{L}(\omega) = 2\sum_{m=1}^{\infty} \frac{\beta_{2m}^{c}}{(2m)!}[(\omega - \omega_{c})^{2m} - \omega_{d}^{2m}] \tag{7}$$

Where $\omega$ is the frequency of signal or idler photons, $\omega_{c} = (\omega_{P1} + \omega_{P2})/2$, $\omega_{d} = (\omega_{P1} - \omega_{P2})/2$, $\beta_{2m}^{c} = (d^{2m}k/d\omega^{2m})_{\omega=\omega_{c}}$ is the 2m-th order dispersion parameter at $\omega_{c}$, and m is any positive integral. An approximation of Eq. (7) up to the fourth order is as follows,

$$\Delta k_{L}(\omega) \approx \beta_{2}^{c}[(\omega - \omega_{c})^{2} - \omega_{d}^{2}] + \frac{\beta_{4}^{c}}{12}[(\omega - \omega_{c})^{4} - \omega_{d}^{4}] \tag{8}$$

Usually, the first term in Eq. (8), namely the second order dispersion term is enough to approximate the linear phase mismatch. However, when the average of the two pump frequencies is close to the frequency corresponding to the zero dispersion wavelength (ZDW) of the waveguide, the contribution of the second order dispersion becomes quite small and the contribution of the fourth order dispersion (presented by the second term in Eq. (8)) should be considered.

For non-degenerate SFWM, the nonlinear phase mismatch in Eq. (5) is [31]

$$\Delta k_{NL} = \gamma(P_{1} + P_{2}) \tag{9}$$

Where $P_{1}$ and $P_{2}$ represent powers of the two pump lights, respectively, and $\gamma$ is the effective nonlinear coefficient of the waveguide at $\omega_{c}$. From Ref [32,33], we derive $\gamma$ as follows,

$$\gamma(\omega) = \frac{\omega n_{2}}{c} \frac{n_{0}^{2} \iint_{R_{core}} |\mathbf{E}(x,y)|^{4} dxdy}{Z_{0}^{2} \left| \iint_{R_{total}} \text{Re}\{\mathbf{E}(x,y) \times \mathbf{H}^{*}(x,y) \cdot \mathbf{e}_{z} dxdy\} \right|^{2}} \tag{10}$$

where $n_{0} = 3.48$ and $n_{2} = (4.5 \pm 1.5) \times 10^{-18} m^{2}W^{-1}$ are the linear and nonlinear refractive index of silicon at 1550 nm, respectively, $c$ is the speed of light in the vacuum, $Z_{0} = \sqrt{\mu_{0}/\varepsilon_{0}} = 377\Omega$ is the free-space wave impedance, $\mathbf{E}(x,y)$ and $\mathbf{H}(x,y)$ are the electric and magnetic field vectors of the waveguide mode, respectively, $\mathbf{e}_{z}$ is the unit vector pointing in positive z-direction, $R_{total}$ denotes the total cross section of the waveguide, while $R_{core}$ represents the core region.

The gain of four wave mixing can be calculated by [31]

$$G = \frac{4\gamma^2 P_1 P_2}{g^2} \sinh^2(gL) \tag{11}$$

Where

$$g = [4\gamma^2 P_1 P_2 - (\Delta k / 2)^2]^{1/2} \tag{12}$$

In Eq. (12), $\Delta k$ can be obtained by substituting Eqs. (8) and (9) into Eq. (5).

For degenerate SFWM pumped by mono-color light with frequency of $\omega_p$ and power of $P$, the nonlinear phase mismatch in Eq. (5) is [31]

$$\Delta k_{NL} = 2\gamma P \tag{13}$$

where $\gamma$ is the effective nonlinear coefficient of the waveguide at $\omega_p$, calculated by Eq.(10). In such case, $\omega_c$ and $\omega_d$ in Eq. (7) become $\omega_P$ and 0, respectively, according to $\omega_{P1} = \omega_{P2} = \omega_P$. Hence, Eq. (7) and (8) can be simplified to

$$\Delta k_L(\omega) = 2\sum_{m=1}^{\infty} \frac{\beta_{2m}}{(2m)!}(\omega - \omega_p)^{2m} \tag{14}$$

and

$$\Delta k_L(\omega) \approx \beta_2(\omega - \omega_p)^2 + \frac{\beta_4}{12}(\omega - \omega_p)^4 \tag{15}$$

respectively, where $\beta_{2m}$ is the *2m*-th order dispersion parameter at $\omega_P$. In this case, the gain of four wave mixing can be calculated by [31]

$$G = \frac{\gamma^2 P^2}{g^2} \sinh^2(gL) \tag{16}$$

where

$$g = [\gamma^2 P^2 - (\Delta k / 2)^2]^{1/2} \tag{17}$$

In Eq. (17), $\Delta k$ can be calculated by substituting Eqs. (13) and (15) into Eq. (5).

According to the semi-classical perspective, signal/idler photons in SFWM are generated by the parametric amplification of the vacuum fluctuation, in which the average value of the signal/idler photon flux within unit frequency is $G$ [34]. Therefore, Eqs. (11) and (16) can be used to calculate the biphoton spectra of non-degenerate and degenerate SFWM processes, respectively.